\documentclass[frontmatterverbose,showkeys,amsmath,amssymb,aps,showpacs,reprint,pre]{revtex4-2}
\usepackage{orcidlink}
\usepackage{amsmath}
\usepackage{amsfonts}
\usepackage{graphicx}
\usepackage{bm}
\usepackage{dcolumn}
\usepackage{bm}
\usepackage[mathlines]{lineno}
\usepackage{hyperref}
\hypersetup{
    colorlinks=true,
    linkcolor=blue,
    filecolor=magenta,      
    urlcolor=blue,
    citecolor=blue,
}

\begin{document}
\title{Particle-Based Framework for Continuous Fields of Coupled Phase Oscillators: Exploring Spontaneous Local Synchronization}
\author{Hugues Berry\orcidlink{0000-0003-3470-683X}}
\email{hugues.berry@inria.fr}
\author{Jan-Michael Rye\orcidlink{0009-0005-0109-6598}}
\email{jan-michael.rye@inria.fr}
\author{Leonardo Trujillo\orcidlink{0000-0001-9995-4135}}%
\email{leonardo.trujillo@inria.fr}
\affiliation{
AIstroSight, \href{https://ror.org/022gakr41}{Inria Lyon Centre}, \href{https://ror.org/01502ca60}{Hospices Civils de Lyon},\href{https://ror.org/029brtt94}{Université Claude Bernard Lyon 1}, 
F-69603 Villeurbanne, France
}
\date{\today}
\begin{abstract}
We introduce a particle-based framework inspired by smoothed particle hydrodynamics (SPH) to simulate the dynamics of a continuous field of coupled phase oscillators. 
This methodology discretizes the spatial domain into particles and employs a smoothing kernel to model non-local interactions, enabling the exploration of how spatial heterogeneities and interaction ranges influence the synchronization and pattern formation of coupled phase oscillators. 
 Notably, we observe the emergence of spatially localized synchronization clusters, providing evidence for spontaneous local synchronization in these systems. This local synchronization refers to the transition from an initially homogeneous state, where no preferred spatial organization exists, to one where structured synchronization patterns emerge due to local interactions. Our results advance the theoretical understanding of spatiotemporal synchronization and demonstrate the utility of SPH-inspired techniques for modeling complex, spatially distributed systems. 
These findings are particularly relevant to applications where spatial interactions drive collective dynamics, such as in neural systems, ecosystems, power grids, social models, chemical oscillators, and climate systems, as well as in condensed matter and collective phenomena involving synchronization.
\end{abstract}
\maketitle

\section{Introduction}
Traditionally, coupled phase oscillators are studied using the Kuramoto model~\cite{kuramoto1975self,kuramoto1984chemical,strogatz2000kuramoto,acebron2005kuramoto}, which approaches synchronization from a purely temporal perspective, without explicitly accounting for spatial influences. However, many physical and biological systems exhibit synchronization patterns that depend not only on temporal dynamics but also on spatial configurations~\cite{pikovsky2001synchronization,cumin2007generalising,breakspear2010generative,cabral2011role,lynn2019physics,anyaeji2021quantitative,blasius1999complex,tsonis2007new,fan2021statistical}.

Such spatio-temporal considerations are particularly important in the study of diverse real-world systems. Examples include brain networks with spatially dependent interactions~\cite{cumin2007generalising,breakspear2010generative,cabral2011role,lynn2019physics}, applications in clinical neuroscience~\cite{anyaeji2021quantitative}, ecosystems shaped by spatial heterogeneity~\cite{blasius1999complex}, and geophysical systems where synchronization is modulated by geographical distances~\cite{tsonis2007new,fan2021statistical}.

Although most studies on spatially distributed oscillators have focused on nonglobal couplings and the impact of network topology on synchronization~\cite{boccaletti2006complex,arenas2008synchronization,rodrigues2016kuramoto}, a significant body of research has focused on chimera states, where synchronized and desynchronized regions coexist~\cite{kuramoto2002coexistence,abrams2004chimera,panaggio2015chimera,parastesh2021chimeras}. 
Despite this, there are still open questions surrounding other synchronization patterns and transitions in systems with continuous spatial dynamics. For instance, how do spatial heterogeneities, such as local variations in coupling strengths, influence the emergence and stability of synchronization patterns? Additionally, what role does the interaction range, local or non-local, play in governing transitions between different synchronization regimes, including the emergence of locally synchronized clusters and global synchronization? Furthermore, how might boundary conditions (e.g. open or periodic) modify synchronized states in these systems?

Building on these open questions, our work investigates a continuous model of coupled phase oscillators that incorporates both spatial and temporal dynamics, aiming to extend the current understanding beyond the traditional emphasis on network topology and discrete couplings. This approach is particularly relevant in fields such as neuroscience, where capturing the interplay between local and non-local interactions is crucial to accurately model brain dynamics~\cite{breakspear2017dynamic,wang2024virtual}. By shifting the focus to spatial domains, our study aims to provide deeper insights into how spatially distributed oscillators achieve synchronization, revealing patterns and transitions that may be overlooked in traditional discrete models.

In this paper, we introduce an alternative approach to the traditional numerical and analytical techniques commonly used to study Kuramoto continuum models, e.g. \cite{strogatz2000kuramoto,kuramoto2002coexistence,abrams2004chimera,acebron2005kuramoto,panaggio2015chimera}. Specifically, we develop a particle-based method inspired by smoothed particle hydrodynamics (SPH)~\cite{gingold1977smoothed,lucy1977numerical,monaghan1992smoothed,liu2003smoothed,liu2015particle} to efficiently simulate spatiotemporal synchronization dynamics. Particle-based methods like SPH offer a flexible, mesh-free framework that can handle complex geometries and dynamic interactions, making them ideal for modeling systems with spatio-temporal dependencies. Originally developed for astrophysics~\cite{gingold1977smoothed,lucy1977numerical,monaghan1992smoothed} and fluid dynamics~\cite{monaghan2012smoothed,ye2019smoothed}, SPH excels in approximating continuous fields using discrete particles, allowing for adaptive resolution. 

Our study reveals several key findings on spatiotemporal synchronization in the particle-based smoothed Kuramoto model. First, we demonstrate the emergence of localized synchronization clusters driven by finite-range interactions, highlighting a spontaneous local synchronization phenomenon. Here, local synchronization refers to the transition from an initially homogeneous state, where no preferred spatial organization exists, to one where structured synchronization patterns emerge due to local interactions. Although the initial conditions are sampled from a uniform distribution, local variations drive self-organization, leading to spontaneous formation of synchronization clusters.
Second, we identify transitions between asynchronous, locally synchronized, and globally synchronized regimes, controlled by the interaction kernel length. Third, we show that the system's dynamics are robust to variations in initial conditions and coupling strength, emphasizing the intrinsic role of spatial organization. 
These results advance our understanding of synchronization patterns in spatially extended systems and provide a new framework for studying self-organization in distributed networks.

\section{Model}
\subsection{\label{sec:spatio_temporal_kuramoto_model}Spatiotemporal Model}
Consider a continuous field of phase oscillators \( \phi(\mathbf{x}, t) \) defined in the spatial domain \( \Omega \subseteq \mathbb{R}^d \) and time \(t\in\mathbb{R}^{+}\). The evolution of the system is governed by the spatio-temporal equation~\cite{kuramoto2002coexistence}:
\begin{multline}
\frac{\partial \phi(\mathbf{x}, t)}{\partial t} = \omega(\mathbf{x}) + \int_{\Omega} G(\mathbf{x} - \mathbf{y}) \\
\times\sin(\phi(\mathbf{y}, t) - \phi(\mathbf{x}, t)) \, d\mathbf{y},
\label{eq:Kuramoto}
\end{multline}
with the initial condition \(\phi(\mathbf{x},t_0)=\phi_0(\mathbf{x})\).

In this initial value problem, we consider oscillators driven by a uniform natural frequency  \(\omega(\mathbf{x}) = \omega\)  for all  \(\mathbf{x} \in \Omega\), allowing us to isolate the effects of spatial interactions and coupling strengths on synchronization patterns. The continuous kernel \(G(\mathbf{x} - \mathbf{y})\) establishes nonlocal coupling between oscillators, with its strength decaying as a function of spatial separation~\cite{kuramoto2002coexistence,abrams2004chimera,panaggio2015chimera}. Such a formulation aligns with neural field models, where neural activity is treated as a continuous spatial field, effectively integrating both local and long-range interactions to reflect physiological observations~\cite{amari1977dynamics,bressloff2011spatiotemporal,cook2022neural}.

To quantify phase coherence across \(\Omega\), we introduce a continuous order parameter, following the approach in Ref.~\cite{kuramoto2002coexistence} and further elaborated in Ref.~\cite{abrams2004chimera}, defined as
\begin{equation}
R(\mathbf{x},t) e^{i \phi(\mathbf{x},t)}  := \frac{1}{|\Omega|} \int_{\Omega} G(\mathbf{x} - \mathbf{y})e^{i \phi(\mathbf{y}, t)} \, d\mathbf{y}.    
\label{eq:order_parameter}
\end{equation}

The order parameter  \(R(\mathbf{x}, t) \in [0, 1]\)  quantifies local phase coherence at a given position \(\mathbf{x}\) within the spatial domain  \(\Omega\), where  \(R \approx 1\)  indicates high synchronization, while  \(R \approx 0\)  corresponds to desynchronization.
When the interaction kernel \( G(\mathbf{x} - \mathbf{y}) = 1 \), all oscillators contribute equally to \( R(\mathbf{x}, t) \), reducing the system to a globally coupled form akin to the canonical Kuramoto model. In contrast, a non-trivial kernel \( G(\mathbf{x} - \mathbf{y}) \neq 1 \) introduces spatial dependence.

\subsection{\label{sec:smoothed_kuramoto_model}Smoothed Kuramoto Model}
Here, we build upon the principles of SPH to develop a particle-based method for modeling a continuous field of coupled phase oscillators. This approach relies on two key features: the particle approximation and the kernel approximation~\cite{liu2003smoothed,liu2015particle}, which enable the transformation of continuous models, such as Eq.~(\ref{eq:Kuramoto}), into a discrete particle-based framework. These approximations not only facilitate the adaptation of continuous models but also preserve the spatially distributed nature of the interactions, ensuring an accurate representation of the system’s phenomenology.

The \textit{particle approximation} discretizes the spatial domain \( \Omega \) into a finite set of particles located at position \( \mathbf{x}_i \) for \( i = 1, 2, \dots, N \). Each particle is assigned a phase \( \psi_i(t) \approx \phi(\mathbf{x}_i, t) \), corresponding to the phase field evaluated at its position, and a natural frequency \( \omega_i \approx \omega(\mathbf{x}_i) \). For simplicity, we assume a uniform natural frequency, such that \( \omega(\mathbf{x}_i) = \omega \) for all \( \mathbf{x}_i \in \Omega \).

The \textit{kernel approximation} replaces the pointwise evaluation in the integral term with a smoothing function  \(W(\mathbf{x} - \mathbf{y}, h)\)~\cite[p.~196]{liu2015particle}. This function is a continuous function with compact support, meaning it is nonzero only within a finite neighborhood defined by the smoothing length \(h\). This modification ensures a localized but smooth approximation of interactions.

The discrete representation of a continuous function \( f(\mathbf{x}) \) in a particle-based framework can be expressed as~\cite{liu2003smoothed,liu2015particle}
\[
f(\mathbf{x}_i) \approx \sum_{j \in \mathcal{N}_i}f(\mathbf{x}_j) \frac{m_j}{\rho_j} W(\mathbf{x}_i - \mathbf{x}_j, h),
\]
where \( \mathcal{N}_i \) denotes the neighborhood of particle \( i \), containing all particles \( j \) within the smoothing radius \( 2h \), \( m_j \) and \( \rho_j \) are the mass and density associated with particle \( j \), and \( W(\mathbf{x}_i - \mathbf{x}_j, h) \) is the smoothing kernel. This kernel ensures smooth interpolation of the function \( f \) based on contributions from neighboring particles, with its influence limited by the smoothing radius \( 2h \).

Using this representation, the continuous integral in Eq.~(\ref{eq:Kuramoto}) can be expressed as a discrete summation over particles. This reformulation not only enables a computationally efficient, particle-based representation of the model but also ensures that the spatial structure of the interactions is preserved. The resulting discrete form is given by (see Appendix \ref{sec:appendixSPH} for details):
\begin{multline}
\int_{\Omega} G(\mathbf{x} - \mathbf{y}) \sin(\phi(\mathbf{y}, t) - \phi(\mathbf{x}, t)) \, d\mathbf{y} \\
\approx \sum_{j \in \mathcal{N}_i} \frac{m_j}{\rho_j} \sin(\psi_j - \psi_i) W(\mathbf{x}_i - \mathbf{x}_j, h).
\nonumber
\end{multline}

Assuming a uniformly distributed particle density, we approximate \( \frac{m_j}{\rho_j} \approx \frac{1}{|\mathcal{N}_j|} \).  Substituting this into the Kuramoto model yields
\begin{equation}
\frac{d \psi_i}{d t} = \omega_i + \frac{1}{|\mathcal{N}_j|} \sum_{j \in \mathcal{N}_i} \sin(\psi_j - \psi _i) W(\mathbf{x}_i - \mathbf{x}_j, h).
\label{eq:smoothed_kuramoto_equation}
\end{equation}

In addition to satisfying compact support and normalization, \( W \) must be symmetric to ensure reciprocal interactions between particles and must decay with distance, assigning greater weight to closer particles to reflect the localized nature of interactions. In our case, to ensure consistency with continuous fields, \( W \) is additionally required to be at least first-order differentiable. For a thorough discussion on the properties of smoothing kernels and their implementation, the reader is referred to \cite{monaghan2012smoothed,liu2003smoothed,liu2015particle,ye2019smoothed}.

\section{\label{sec:methods}Methods}
The simulations are carried out within a square spatial domain $\Omega = L \times L$. Each particle is assigned a natural frequency fixed at $\omega = \pi / 10$ for all particles.
Unless otherwise specified, the initial positions of the particles are sampled from a 2D uniform distribution across the domain. Likewise, the initial phases are independently sampled from a uniform distribution over the interval $[0, 2\pi]$.

The time evolution of the system is obtained by numerically solving Eq.~(\ref{eq:smoothed_kuramoto_equation}) using a fourth-order Runge-Kutta integration scheme. The kernel length \(h\) is varied to explore different synchronization regimes.

To quantify the level of synchronization, we approximate the local order parameter of Eq.~(\ref{eq:order_parameter}) by a discrete local estimate:
\[
\hat{R}(\mathbf{x}) = \frac{1}{|\mathcal{B}(\mathbf{x},\xi)|} \left| \sum_{\mathbf{x}' \in \mathcal{B}(\mathbf{x},\xi)} e^{i \varphi(\mathbf{x}')} \right|,
\]
where \(\mathcal{B}(\mathbf{x},\xi)\) represents the circular neighborhood of radius \(\xi\) centered at \(\mathbf{x}\), \(|\mathcal{B}(\mathbf{x},\xi)|\) is the number of particles within that neighborhood, and \(e^{i \varphi(\mathbf{x}')}\) denotes a unit vector in the complex plane at angle \(\varphi(\mathbf{x}')\). The value of \(\hat{R}(\mathbf{x})\) ranges from $0$, indicating complete desynchronization, to $1$, corresponding to full synchronization. This parameter captures the local coherence of phases and complements the global dynamics described by the kernel interactions.

To further characterize the spatial variability in synchronization, we introduce the \textit{contrast}
\[
C := \frac{\sigma(\hat{R}(\mathbf{x}))}{\langle \hat{R}(\mathbf{x}) \rangle},
\]
where \(\sigma(\hat{R}(\mathbf{x}))\) is the standard deviation and \(\langle \hat{R}(\mathbf{x}) \rangle\) the mean of \(\hat{R}(\mathbf{x})\) across the domain. High \(C\) values are obteind for system where the phases are fully randomized spatially, whereas \(C\) vanishes for uniform synchronization across the domain. In between, spatial heterogeneity and spatially-defined synchronization clusters, manifest as low but non zero values of \(C\). Together, \(R(\mathbf{x})\) and \(C\) provide a quantitative framework for analyzing the emergence and spatial organization of synchronization patterns.

To accurately capture spatial interactions, we employ a two-dimensional piecewise cubic spline kernel. This kernel is widely used in SPH applications due to its computational efficiency and smoothness~\cite[p. 90]{liu2003smoothed}. The kernel is defined as
\[
W(r, h) = \frac{{200}}{7\pi h^2}
\begin{cases}  
 \left( 1 - \frac{3}{2}q^2 + \frac{3}{4}q^3 \right), & 0 \leq q < 1, \\[8pt]
\left( \frac{1}{4}(2 - q)^3 \right), & 1 \leq q < 2, \\[8pt]
0, & q \geq 2,
\end{cases}
\]
where \( r = |\mathbf{x}_i - \mathbf{x}_j| \) and \( q = r/h \). The kernel has compact support, meaning \( W(r, h) \neq 0 \) only for \( r < 2h \), which restricts interactions to neighboring particles and ensures efficient computation. The neighborhood of each particle \(i\) satisfies \( \mathcal{N}_i \leq 2h\), limiting interactions to a finite region around each particle. This localized interaction is central to the emergence of spatial synchronization domains observed in the results.

To compute \(\hat{R}(\mathbf{x})\), the domain \(\Omega\) is discretized into a square lattice with a spacing of \(L/40\), where \(\hat{R}(\mathbf{x})\) is evaluated at each lattice site. To account for variations in particle density, the neighborhood radius is defined as \(\xi = \beta L / N\), with \(\beta = 150\). For instance, when \(L = 20\) and \(N = 2,000\), this results in \(\xi = 1.5\). This setup ensures that interactions are captured efficiently while maintaining spatial resolution, complementing the kernel's compact support. 

To estimate the instantaneous frequency and phase offset of each particle $i$ at time $t$, we assume that the time evolution of $i$'s phase is linear over a short sliding temporal window of duration $T$, i.e. $\psi_i(t)=\hat{\eta}_it + \hat{\gamma}_i$ where $\hat{\eta}_i$ and $\hat{\gamma}_i$ are the instantaneous frequency and the phase offset, respectively. We estimate the instantaneous frequency as the median of the numerical derivative of $\psi_i$ over the time period, i.e. $\hat{\eta}_i(t)=\mathrm{med}_{\tau\in [-T/2,T/2]}\left((\psi_i(t+\tau+dt)-\psi_i(t+\tau))/dt\right)$ and the offset as $\hat{\gamma}_i(t)=\mathrm{med}_{\tau\in [-T/2,T/2]}\left(\psi_i(t+\tau)-\hat{\eta}_i(t+\tau)\right)$.

The source code of all models and stimulations used in the present paper can be found in \url{https://gitlab.inria.fr/aistrosight/kuramoto_sph}.

\section{\label{sec:results}Results}
\subsection{Multicluster Dynamics and Local Synchronization}

\begin{figure}[ht!]
    \includegraphics[scale=1]{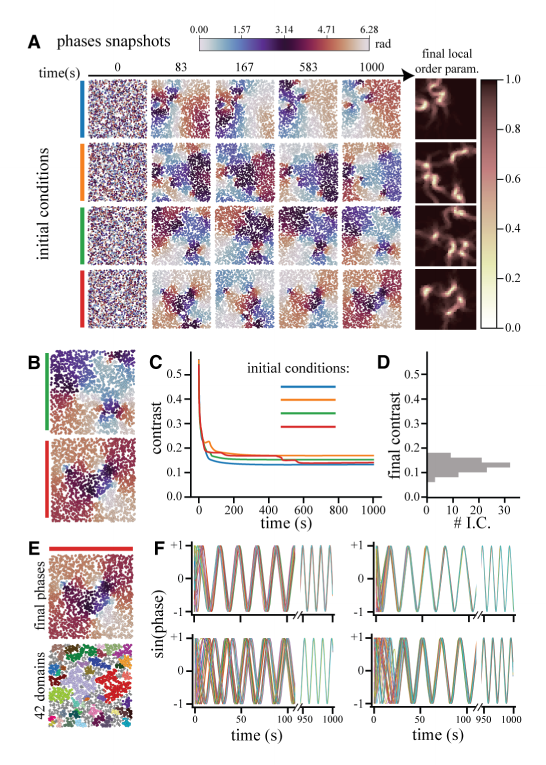}
    \caption{
    {\bf Illustration of local synchronization in the smoothed Kuramoto model}. Snapshots of the time evolution of the particle phases for 4 different initial conditions (IC) (\textbf{A}, one IC per row). The particle positions are sampled from a uniform distribution over the space $\Omega=L\times L$. Each particle $i$ is rendered by a full circle of radius $h$ centered at its position $\mathbf{x}_i$ and colorcoded according to its phase $\psi_i(t)$ (horizontal colorbar on top of the figure). The rightmost column shows a map of the local order parameter $\hat{R}(\mathbf{x})$ at time $t=1,000$ for every IC, colorcoded according to the vertical colorbar. A zoom on the snapshots at $t=1,000$ for the green and red IC is shown in (\textbf{B}). The evolution with time of the  contrast $C$ for each IC is shown in (\textbf{C}) keeping the same colorcode as before. A histogram of the contrast at $t=1,000$ for 100 IC is provided in (\textbf{D}) in vertical format. We then isolated the resulting locally synchronized spatial clusters at $t=1,000$. 
    The result is illustrated for the red IC in (\textbf{E}), with the particles colorcoded according to their phase (top) or spatial cluster (bottom, one color per spatial cluster). Gray crosses indicate particles that do not belong to any cluster (``noise'' samples in DBSCAN). We illustrate in (\textbf{F}) the time evolution of the phases of 50 randomly chosen particles in four randomly chosen clusters, evincing progressive synchronization locally in the clusters.  Common parameters: $N=2,000$ particles, domain size $L=20$, kernel length $h=0.5$.
    }
    \label{fig:wide_figure}    
\end{figure}

\paragraph{Local Synchronization Dynamics and Intensity.}
The smoothed Kuramoto model exhibits spontaneous self-organization into spatial synchronization domains, driven by local interactions. Initially, particles are globally desynchronized, with their positions \({\bf x}_i\) and phases \(\psi_i\) sampled from uniform distributions over the domain \(\Omega\) and \([0, 2\pi]\), respectively. Over time, localized clusters of synchronized oscillators emerge, forming stable spatial patterns that persist across the domain (Fig.~\ref{fig:wide_figure}A, left panel).

The right panel of Fig.~\ref{fig:wide_figure}A shows the spatial distribution of \(\hat{R}(\mathbf{x})\) at the final state (\(t = 1,000\)). The results reveal large regions with high \(\hat{R}(\mathbf{x})\), indicating strong phase coherence, alongside smaller regions of partial synchronization. These patterns emerge from the interplay between local coupling and spatial heterogeneity, leading to a stable configuration where synchronization domains are well-defined and persist over time. This self-organization illustrates the system's ability to form coherent structures, driven by local interactions.

\paragraph{Contrast as a Measure of Spatial Variability.}
The zoomed views in Fig.~\ref{fig:wide_figure}B for the green and red initial conditions provide a high-resolution depiction of the spatial phase distribution at \(t = 1,000\). These snapshots highlight the well-defined synchronization clusters that emerge from the initial random configurations, illustrating the local coherence achieved by the system despite the spatial heterogeneity. While the initial distribution of particles is uniform in expectation, local variations in density and interaction strength naturally lead to structured synchronization patterns. This transition marks a form of spontaneous local synchronization, where the system evolves from a statistically homogeneous state to one with emergent spatial organization.
Figure \ref{fig:wide_figure}C  provides a detailed analysis of \(C\) in different contexts. The time evolution of \(C\) for four representative initial conditions is shown in Fig.~\ref{fig:wide_figure}C, using the same color code as the snapshots in Fig.~\ref{fig:wide_figure}A. Initially, \(C\) is high, reflecting the desynchronized state where the spatial variability in \(\hat{R}(\mathbf{x}\)) is significant. As the system evolves, \(C\) decreases and stabilizes around \(t \approx 200\) to lower but non-zero values (0.1 to 0.2), indicating the formation of stable synchronization clusters. These clusters persist until the end of the simulation at \(t = 1,000\), indicating that the system has reached a steady-state configuration of well-defined synchronization domains.

To provide a broader statistical perspective, Fig.~\ref{fig:wide_figure}{D} presents a histogram of \(C\) at \(t = 1,000\) for 100 independent initial conditions. The distribution demonstrates variability in the final degree of spatial heterogeneity across initial conditions. Higher  \(C\)  values indicate greater spatial variation in synchronization, where distinct localized clusters form, while lower  \(C\)  values correspond to more uniform synchronization across the domain. On average, we find  \(\langle C \rangle = 0.127\) with a standard deviation of \(\sigma_C = 0.024\), highlighting the range of possible final configurations.

These results underscore the interplay between initial conditions and system dynamics in shaping spatial synchronization patterns. Non-zero  \(C\)  values at  \(t = 1,000\)  correspond to scenarios in which distinct spatially localized domains of synchronized oscillators persist, while vanishing values indicate transitions toward more homogeneous synchronization across the system.
This highlights the stochastic nature of the system dynamics and the sensitivity to initial conditions.

\paragraph{Identification and Dynamics of Local Synchronization Domains.}
To further investigate the emergence of spatial organization at \(t = 1,000\), we applied the DBSCAN clustering algorithm from the scikit-learn library~\cite{scikit-learn}\footnote{The DBSCAN algorithm was used with the parameters \texttt{eps} = 0.045 and \texttt{min\_samples} = 10, unless otherwise stated.}. This method enabled the identification of locally synchronized spatial domains by grouping particles based on their spatial proximity and phase coherence. The results for the red initial condition are presented in Fig.~\ref{fig:wide_figure}E. In the upper panel, particles are color-coded according to their phase, revealing regions of coherent phase alignment, while the lower panel assigns unique colors to each spatial domain identified by the clustering algorithm. Particles not assigned to any cluster, identified as noise by DBSCAN, are shown as gray crosses. These unsynchronized particles represent isolated dynamics or transitions between two synchronized domains, further emphasizing the spatially localized nature of synchronization.

The clustering results underscore the system's ability to self-organize into distinct domains where local interactions dominate over global coupling, leading to phase coherence within spatially restricted regions. This phenomenon reflects a form of spontaneous local synchronization, where an initially unstructured distribution of phases and particle positions gives rise to synchronized spatial patterns mediated by finite-range coupling. While the spatial distribution of particles exhibits some inherent heterogeneity, this heterogeneity only partially determines the final synchronization domains, emphasizing the emergent nature of these structures.

To explore the temporal evolution of synchronization within these domains, we analyzed the phase trajectories of 50 randomly chosen particles from four randomly chosen clusters. The results, shown in Fig.~\ref{fig:wide_figure}F, reveal the progressive alignment of phases within each domain. This convergence highlights the local evolution of phase synchronization within clusters, despite the absence of global synchronization.

\begin{figure}[ht!]
    \includegraphics[scale=1.2]{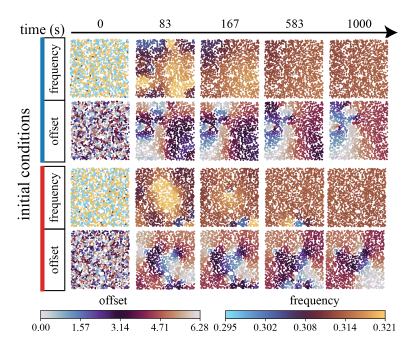}
    \caption{
    {\bf Spatial clusters in the smoothed Kuramoto model are defined by common offset}. Snapshots of the time evolution of the particle instantaneous frequency (top rows) and offset (bottom rows) for the blue and red initial conditions (IC) of Fig.~\ref{fig:wide_figure}. The colorbars for both observables are shown below the main panel. See the text for the definitions of the instantaneous frequancy and offset. Common parameters: $N=2,000$ particles, domain size $L=20$, kernel length $h=0.5$.
    }
    \label{fig:freq_offset}    
\end{figure}

To further advance the analysis, we estimated the medium instantaneous frequency $\hat{\eta}_i(t)$ and phase offset $\hat{\gamma}_i(t)$ of each particle $i$ at time t (see Methods section~\ref{sec:methods}) for two ICs. In  Fig.~\ref{fig:freq_offset} for each IC, the snapshots of the instantaneous frequency reveal that the particles progressively assume the same instantaneous frequency (top rows), whereas the instantaneous offsets (bottom rows) change from one cluster to the next. In other words, with time, all the particles converge towards a common global frequency. The clusters are defined by a specific value of the offset, i.e. each particle within a given cluster adopts the same offset, which is different from that of the other clusters.

These dynamics are consistent with the expected behavior in systems where interactions exhibit spatial structure, such as certain neural circuits or coupled oscillator arrays.

\subsection{Transitions Between Synchronization Regimes}
\begin{figure}[ht!]
    \centering
    \includegraphics[scale=1.2]{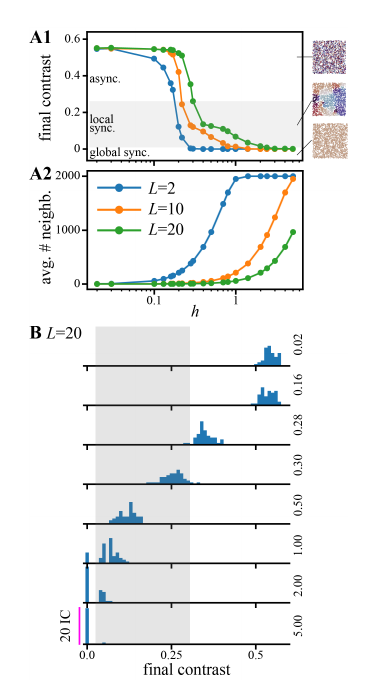}
    \caption{
    {\bf Transitions between globally asynchronized, locally synchronized and globally synchronized regimes}. We evaluated the final contrast $C$ (at time $t=1,000$, $\mathbf{A1}$) and the average number of neighbors within distance $2h$ ($\mathbf{A2}$) for a score of $h$ values ($x$-axis) and space sizes $L$ (colorcode in panel \textbf{A2}). Shown are averages over $20$ simulations with different IC. The shaded area in \textbf{A1} marks the region where local synchronization occurs in restricted spatial domains. Above it, the particles remain globally and locally unsynchronized, while they are all completely synchronized below it. Illustrations for these three regimes are given as vignette snapshots of the phases, at the right of the upper panel. The distributions of the final contrasts over 20 IC and for a selection of $h$ values (vertical labels) for $L=20$ are shown in ($\mathbf{B})$. The pink bar at the bottom represents a count of 20 IC and the shaded area marks the local synchronization region. In all panels, $N=2,000$ particles.
    }
    \label{fig:plot_along_h}    
\end{figure}

To understand the emergence of distinct synchronization regimes, we systematically evaluated the behavior of the smoothed Kuramoto model Eq.~(\ref{eq:smoothed_kuramoto_equation}) across a range of kernel lengths \(h\). The results, summarized in Fig.~\ref{fig:plot_along_h}, reveal transitions between globally asynchronous, locally synchronized, and globally synchronized states as a function of \(h\). These transitions were characterized by the final contrast \(C\), and the average number of neighbors within a distance \(2h\).

\paragraph{Final Contrast as an Indicator of Synchronization.}
Figure~\ref{fig:plot_along_h}A1 reveals three distinct synchronization regimes as a function of the interaction range \(h\), displayed on a logarithmic scale. The contrast \(C\)  and the spatial phase distributions shown in the panels to the right illustrate the corresponding synchronization patterns:
\begin{enumerate}
    \item \textit{Desynchronized regime} (\( h \lesssim 0.1 \)): At small interaction ranges, the contrast  \(C\)  is high (\( C \gtrsim 0.3 \)), reflecting significant spatial heterogeneity due to random fluctuations, rather than structured synchronization clusters. The corresponding panel displays a uniformly disordered phase distribution, indicating that the limited interaction range inhibits the emergence of coherent synchronization domains.
    
    \item \textit{Clustered regime}: The shaded region highlights the regime of local synchronization. This marks the transition to a regime where localized synchronization clusters emerge. The panel associated with this regime shows distinct clusters of coherent phases, highlighting the onset of {\it local synchronization} driven by finite-range interactions. The value of $h$ at which the clustered regime emerges depends $L$, ranging from $\approx 0.2$ to $\approx 0.4$ when $L$ increases tenfold from $2$ to $20$.
    
    \item \textit{Globally synchronized regime}: At larger interaction ranges, the contrast \(C\) vanishes, indicating global synchronization. The final panel illustrates a nearly uniform phase distribution, characteristic of long-range interactions dominating the system and aligning the oscillators across the entire domain. Here again, the exact value of $h$ that marks the entry into the globally synchronized regime is highly dependent on the domain size $L$: around 0.2 and 1.0 for $L=2$ and $10$, respectively, but around $3$ for $L=20$.  
\end{enumerate}

The logarithmic scale of $h$ emphasizes the rapid transitions between these regimes. At small $h$, interactions are too localized to induce coherence, while at large $h$, interactions span the domain, homogenizing the system. The intermediate clustered regime represents a balance between local coherence and global disorder, with \(C\) serving as an effective metric for capturing these transitions. 
For small domain sizes (\( L = 2 \)), the transition from the asynchronous to the globally synchronous regime is very steep, restricting the locally synchronous clustered regime to a narrow range of  \(h\)-values (\( h \in [0.2, 0.3]\), roughly). In these cases, the pre-existing spatial heterogeneity in particle distribution plays a crucial role in determining synchronization patterns, as finite-range interactions primarily reinforce local density variations. Thus, while local synchronization still occurs in the sense that spatial patterns emerge without an explicit external bias, it is strongly mediated by the initial particle distribution and the scale of interaction.
However, for larger system sizes, the transition starts to deviate from a very steep original decay to a much slower, nearly linear decay when \(C\) reaches $\approx 0.2$.
As a result, the range of  \(h\)-values for which the locally synchronized regime is observed becomes much larger with increasing domain size. 

\paragraph{Neighbor Connectivity.}
Figure~\ref{fig:plot_along_h}A2 depicts the average number of neighbors within a radius of \(2h\). The results show a monotonic increase in connectivity as $h$ increases, highlighting the enhanced coupling strength when the interaction range becomes comparable to the system size. In the globally synchronized regime, the connectivity reaches its maximum; i.e. all particles are coupled to each others, facilitating complete phase coherence across the entire system. Note, however, that no simple relation exists between the transition values of  h  and the number of direct neighbors.
However, it is clear from the figure that the dynamics switches from the asynchronous to the locally synchronous clustered regime at $h$ values for which the average particle is connected to only a small number of neighbors.
Transition to the final globally synchronized regime takes place for much larger neighborhoods, but this still occurs well before full connectivity.

\paragraph{Statistical Distributions of Contrast.}
To capture the variability in synchronization dynamics, Fig.~\ref{fig:plot_along_h}B presents the distribution of the final contrast \(C\) over 20 independent initial conditions for selected \(h\) values. 
The broad distributions of  C  in the local synchronization regime suggest a strong dependence on initial conditions, particularly on the spatial distribution of particles. While stochastic phase assignments contribute to variability, further analysis is needed to distinguish the respective roles of initial positions and phase distributions in determining final synchronization patterns.
By contrast, the globally asynchronous and globally synchronized regimes exhibit narrow distributions, highlighting the uniformity of system-wide behaviors in these states. 

\subsection{Synchronization Dynamics in Clustered Particle Configurations}
\begin{figure}[ht!]
    \centering
    \includegraphics[scale=1.2]{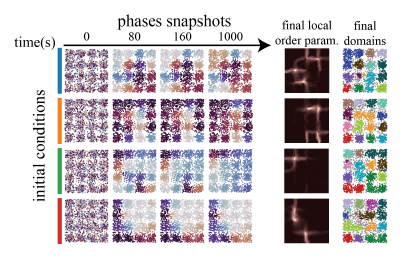}
    \caption{
    {\bf The smoothed Kuramoto model with clustered particles}. Here the positions of the particles were choosen according to a combination of 16 two-dimensional Gaussians, with centers regularly spread over space in a $4\times4$ grid with a variance randomly chosen in the interval $[1.0,3.0]$. The time evolution for 4 IC is illustrated with snapshots of the phases (columns 1-4), maps of the local order parameter $\hat{R}(\mathbf{x})$  (col. 5) or the local domains (col.6) at final time. All other parameters and colorbars were as in Fig.~\ref{fig:wide_figure} except for the particle positions and the radius of the snapshot symbols that where set to $h/6$ here to exhibit the clustering of the positions. Common parameters: $N=2000$ particles, domain size $L=20$, kernel length $h=1.0$.
    }
    \label{fig:clustered}    
\end{figure}

To investigate how the spatial arrangement of particles influences synchronization dynamics, we modified the initial positions of particles to follow a clustered distribution. Specifically, particle positions were sampled from a mixture of two-dimensional Gaussians. 

\paragraph{Temporal Evolution of Phases and Spatial Order.}
The first four columns of Fig.~\ref{fig:clustered} show snapshots of the phase distributions at four different time points (\(t = 0, 80, 160, 1000\)) for four independent initial conditions (ICs). Initially, the phases are randomly distributed, reflecting the disordered nature of the ICs. As time progresses, synchronization domains emerge locally within clusters of particles, with coherent regions becoming more prominent at later times. By \(t = 1,000\), the phases within clusters exhibit substantial alignment, although the global phase coherence remains limited due to the spatial separation between clusters.

\paragraph{Local Order Parameter Maps.}
The fifth column of Fig.~\ref{fig:clustered} presents the spatial distribution of the local order parameter \(\hat{R}(\mathbf{x})\) at \(t = 1,000\). These maps reveal high \(\hat{R}(\mathbf{x})\) values within clusters, indicating strong local synchronization. In contrast, the inter-cluster regions show lower \(\hat{R}(\mathbf{x})\), reflecting weaker interactions between spatially distant clusters. This pattern underscores the role of particle clustering in restricting synchronization to localized domains and highlights that the effects observed with random particle positions primarily arise from their heterogeneous density distributions.

\paragraph{Identification of Local Synchronization Domains.}
The final column of Fig.~\ref{fig:clustered} illustrates the spatial domains identified at \(t = 1,000\). The results show distinct synchronization domains corresponding to the Gaussian clusters in the particle configuration. Each domain exhibits internally coherent phase dynamics, while particles in regions of low density or outside the main clusters are labeled as noise.

\subsection{Spontaneous Local Synchronization with Periodic Boundary Conditions}
\begin{figure}[ht!]
    \centering
    \includegraphics[scale=0.9]{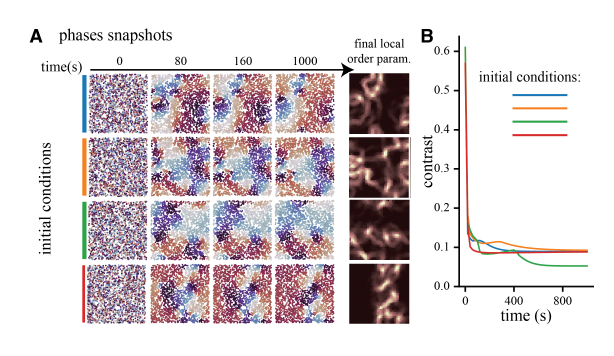}
    \caption{
    {\bf Local synchronization with periodic boundary conditions}. Simulations were prepared with the same parameters and initial conditions as Fig.~\ref{fig:wide_figure} but with inter-particle distances measured under periodic boundary conditions. Snapshots of the time evolution of the phases and maps of the local order parameter at final time ($\mathbf{A}$) and evolution with time of the contrast ($\mathbf{B}$) are shown using the same colorbars and conventions as Fig.~\ref{fig:wide_figure}. Common parameters: $N=2000$ particles, domain size $L = 20$, kernel length $h=0.5$.
    }
    \label{fig:periodic}    
\end{figure}

To explore the influence of boundary conditions on the synchronization dynamics of the smoothed Kuramoto model, we repeated simulations with periodic boundary conditions. This modification ensures that the distances between particles are calculated as the shortest path across the toroidal domain, removing edge effects present in the fixed-boundary setup.  This adjustment allows for a more intrinsic exploration of the self-organization phenomena in spatially extended systems. The results, presented in Fig.~\ref{fig:periodic}, illustrate the dynamics of phase synchronization and the evolution of spatial coherence under periodic boundaries.

\paragraph{Temporal Evolution of Phases and Spatial Patterns.}
Figure~\ref{fig:periodic}A shows snapshots of the particle phases at different time points (\(t = 0, 80, 160, 1000\)) along with maps of the local order parameter \(\hat{R}(\mathbf{x})\) at \(t = 1,000\). The periodic boundary conditions maintain the spatial continuity of the system, allowing synchronization domains to form without interruption at the edges. The snapshots reveal the emergence of localized synchronization patterns, similar to those observed with fixed boundaries. 
At \(t = 1000\), the \(\hat{R}(\mathbf{x})\) maps indicate high local synchronization within well-defined regions, consistent with the local synchronization phenomenon observed in the original setup.

\paragraph{Evolution of Contrast.}
Figure~\ref{fig:periodic}B tracks the time evolution of the contrast \(C\), a measure of spatial variability in the local order parameter \(\hat{R}(\mathbf{x})\). As in the fixed-boundary case, \(C\) initially decreases, reaching a steady state around \(t \approx 200\). This suggests that while periodic boundaries alter the spatial structure of synchronization clusters, they do not fundamentally change the self-organization process.

The smoother patterns observed in the periodic setup show the adaptability of the model to various boundary conditions, offering a closer approximation to real-world systems operating in unbounded or continuous spaces.

\section{\label{sec:discussion}Discussion}
The results presented in this study demonstrate how finite-range interactions induce localized synchronization and spatial clustering patterns in spatially distributed systems. The transition from globally asynchronous states to localized synchronization exemplifies spontaneous local synchronization, a phenomenon analogous to phase transitions in thermodynamic systems~\cite{kuramoto1984chemical,strogatz2001exploring,gupta2014kuramoto}. As interactions evolve, finite-range coupling disrupts global symmetry, leading to the emergence of synchronization clusters~\cite{sakaguchi1986soluble}. This behavior is governed by the parameter  \(h\), which determines the range of interactions relative to the domain size. However, the effect of  \(h\)  is also strongly influenced by the spatial distribution of particles, as local density variations shape the effective coupling landscape. The observed variability in regime transitions across different initial conditions suggests that  \(h\)’s impact is not uniform but depends on the interplay between interaction range and initial spatial heterogeneity. A more detailed analysis of this dependence could provide deeper insight into the robustness and predictability of synchronization transitions.
At intermediate values of \( h \), the system exhibits localized synchronization, representing a regime distinct from both globally asynchronous and globally synchronized states. This local synchronization regime reflects a balance between local coherence and global disorder, driven by the finite range of interactions~\cite{abrams2004chimera,omel2013coherence}.

However, it is important to clarify the nature of the symmetry that is being broken. In our model, the initial distribution of particles is statistically homogeneous, meaning that there is no preferred spatial organization beyond the expected uniformity of particle positions. If the system retained this statistical symmetry, we would expect either complete global desynchronization or homogeneous synchronization. The emergence of structured, localized synchronization clusters represents a spontaneous departure from this initial homogeneity, indicating a form of self-organization akin to pattern formation in reaction-diffusion systems and phase transitions in condensed matter physics. In this sense, local synchronization does not imply the loss of an explicit geometric symmetry, but rather the emergence of a structured state from an initially unstructured one.

In this report, we have chosen to work with constant particle counts and variable domain sizes. This means that particle densities vary with the domain size, thus modifying the effective interaction structure. To isolate the role of the domain size \(L\) independently from density effects, future work should explore variations in domain size while keeping the global particle density approximately constant. 

The synchronization patterns observed in this model reflect key principles of self-organization across diverse physical and biological systems. In neural networks, for instance, localized synchronization clusters resemble modularity, where brain regions synchronize locally while maintaining partial independence, enabling functionalities such as sensory processing and memory consolidation~\cite{bassett2017network,breakspear2017dynamic}. Similarly, in ecological systems, spatial clustering mirrors synchronization within geographically constrained subpopulations, driven by proximity-based interactions\cite{blasius1999complex}. In physical systems, the emergence of synchronization domains bears resemblance to coherent structures such as vortices in fluid dynamics~\cite{kuramoto1984chemical,holmes2012turbulence}. These parallels highlight the versatility of the particle-based smoothed Kuramoto model in capturing spatially distributed interactions across systems governed by local and nonlocal couplings.

Unlike traditional Kuramoto models on networks, which emphasize discrete topologies with fixed adjacency relationships~\cite{boccaletti2006complex,arenas2008synchronization,rodrigues2016kuramoto}, the particle-based framework introduced here approximates a continuous field through spatially defined interactions~\cite{acebron2005kuramoto,montbrio2015macroscopic}.
By modeling interactions as spatially distributed fields, this approach captures phenomena such as spatial clustering driven by finite-range coupling, which are often overlooked in discrete models. This makes it particularly advantageous for studying systems where spatial structure plays a critical role, including neural fields~\cite{byrne2020next}, ecological distributions~\cite{hastings2012encyclopedia}, and geophysical dynamics~\cite{tsonis2007new,fan2021statistical}.

Future extensions of this framework could explore intrinsic heterogeneities, such as non-uniform oscillator frequencies~\cite{pazo2005thermodynamic}, or introduce stochastic effects to capture noise-driven dynamics~\cite{teramae2004robustness,goldobin2005synchronization}. Additionally, expanding the model to three-dimensional domains would allow for the investigation of more complex spatial interactions, with potential applications in fluid dynamics and 3D neural systems. The observed robustness of the synchronization patterns, combined with the flexibility of the particle-based approach, underscores the potential of this methodology to advance our understanding of self-organization and local synchronization phenomena across a wide range of disciplines.

\begin{acknowledgments}
This work benefited from the support of the project EngFlea (Grant No. \href{https://anr.fr/Project-ANR-21-CE16-0022}{ANR Project ANR-21-CE16-0022}) of the French National Research Agency (ANR).  
The authors thank the \href{https://www.grid5000.fr}{Grid5000} testbed, supported by a scientific interest group hosted by Inria and including CNRS, RENATER and several Universities as well as other organizations, for computational support.
\end{acknowledgments}

\appendix
\section{SPH-Based Approach}
\label{sec:appendixSPH}
This appendix provides additional explanations regarding the particle-based (SPH) formulation used throughout the manuscript. The purpose is to clarify how continuous integral terms are replaced by discrete summations and how the partial differential equation (PDE) in \(\phi(\mathbf{x},t)\) is ultimately turned into a system of ordinary differential equations (ODEs) for \(\varphi_i(t)\), without delving into details of Riemann sums or other numerical solvers.

SPH introduces a smoothing kernel \(W(\mathbf{x}-\mathbf{y},h)\) that approximates the Dirac delta,
\[
\delta(\mathbf{x}-\mathbf{y})
\;\approx\;
W(\mathbf{x}-\mathbf{y},h),
\]
where \(h\) is the \emph{smoothing length}. Let \(\Omega \subset \mathbb{R}^d\) be the spatial domain under consideration. In practice, the kernel must satisfy the normalization condition
\[
\int_{\Omega} W(\mathbf{x}-\mathbf{y},h)\,d\mathbf{y} 
\;\approx\; 1
\quad
\forall\;\mathbf{x},\,\mathbf{y} \in \Omega,
\]
along with a locality (or ``Dirac convergence'') condition such as
\[
\lim_{h\to\infty} W(\mathbf{x}-\mathbf{y},h)= \delta(\mathbf{x}-\mathbf{y}).
\]
Here, the family of functions \(W(\mathbf{x}, h)\) \emph{converges to} \(\delta(\mathbf{x})\) \textit{in the sense of distributions} (or \emph{weak sense}). That is, for any test function \(\varphi(\mathbf{x})\) (smooth, typically with compact support), the following holds
\[
\lim_{h \to 0}
\int_{\Omega}
W\bigl(\mathbf{x}-\mathbf{y}, h\bigr)\,\varphi(\mathbf{x})
\;d\mathbf{x}
\;=\;
\varphi(\mathbf{y}),
\]
which is precisely the definition of \(\delta(\mathbf{x})\) as a distribution.

In addition, \(W\) is predominantly concentrated within a ball of approximate radius \(2h\) around \(\mathbf{x}\). Next, one discretizes \(\Omega\) by introducing \(N\) particles, each labeled by an index \(j\). Particle \(j\) has position \(\mathbf{x}_j\), mass \(m_j\), and density \(\rho_j\). Continuous integrals are then replaced by discrete sums. For a function \(f(\mathbf{y})\), the SPH approach yields
\[
\int_{\Omega} f(\mathbf{y})\,d\mathbf{y}
\;\;\approx\;\;
\sum_{j}
f(\mathbf{x}_j)\,\frac{m_j}{\rho_j}.
\]
When localized smoothing is desired, one incorporates the kernel:
\[
\int_{\Omega} f(\mathbf{y})\,W(\mathbf{x}-\mathbf{y},\,h)\,d\mathbf{y}
\;\;\approx\;\;
\sum_{j}
f(\mathbf{x}_j)\,\frac{m_j}{\rho_j}\,
W\bigl(\mathbf{x}-\mathbf{x}_j,\,h\bigr).
\]

Now let's consider the integral
\[
I(\mathbf{x}, t)
\;=\;
\int_{\Omega} 
G\bigl(\mathbf{x}-\mathbf{y}\bigr)\,\sin\!\Bigl(\phi(\mathbf{y},t)-\phi(\mathbf{x},t)\Bigr)\,d\mathbf{y},
\]
where \(G(\mathbf{x}-\mathbf{y})\) is a weighting (or coupling) function, and \(\sin\!\bigl[\phi(\mathbf{y},t)-\phi(\mathbf{x},t)\bigr]\) is the integrand in \(\mathbf{y}\). By identifying \(G\) with a suitable kernel \(W\), we can approximate \(I(\mathbf{x}, t)\) through a discrete sum over neighboring particles. Specifically, setting \(\mathbf{x}=\mathbf{x}_i\) as the position of particle \(i\) and denoting the neighbor set by \(\mathcal{N}_i\), we write
\[
I(\mathbf{x}_i,t)
\;\approx\;
\sum_{j \,\in\, \mathcal{N}_i}
\frac{m_j}{\rho_j}\,
\sin\!\bigl[\phi(\mathbf{x}_j,t)-\phi(\mathbf{x}_i,t)\bigr]\,
W\bigl(\mathbf{x}_i-\mathbf{x}_j,\,h\bigr).
\]
Introducing \(\psi_i(t)=\phi(\mathbf{x}_i,t)\) and \(\psi_j(t)=\phi(\mathbf{x}_j,t)\), yields the simpler expression
\begin{multline}
\int_{\Omega} G\bigl(\mathbf{x}_i - \mathbf{y}\bigr)\,\sin\!\Bigl(\phi(\mathbf{y},t)-\phi(\mathbf{x}_i,t)\Bigr)\,d\mathbf{y} 
\\
\approx
\sum_{j \,\in\, \mathcal{N}_i}
\frac{m_j}{\rho_j}\,
\sin\!\bigl(\psi_j - \psi_i\bigr)\,
W\bigl(\mathbf{x}_i-\mathbf{x}_j,h\bigr),
\nonumber
\end{multline}
transforming the nonlocal operator into a discrete sum over neighbors.
See \cite[Chapters~2–3]{liu2003smoothed} where the integral approximation and SPH interpolation concepts are laid out in detail to justify replacing continuous integrals by discrete summations.

Once all spatial (or nonlocal) terms are replaced by SPH-based summations, one typically rewrites
\(\tfrac{\partial \phi}{\partial t}\bigl|_{\mathbf{x}=\mathbf{x}_i}\)
as
\(\tfrac{d\psi_i}{dt}\).
Strictly speaking, \(\tfrac{\partial \phi}{\partial t}\bigl|_{\mathbf{x}=\mathbf{x}_i}\) is not interpolated like a spatial function; instead, it becomes an ordinary derivative \(\tfrac{d\psi_i}{dt}\). This occurs because the SPH formulation has already removed the continuous spatial dependence in favor of a discrete set of particles.
Consequently, the time evolution of \(\psi_i\) follows from solving the ODE
\[
\frac{d\psi_i}{dt}
\;=\;
F\bigl(\{\psi_j\}\bigr),
\]
where \(F\) is the discrete right-hand side obtained from the original PDE (now expressed in terms of sums over neighbor particles). One then integrates \(\psi_i(t)\) forward in time using a standard ODE method (e.g. explicit Euler, Runge-Kutta). In other words, after discretizing the spatial operators, the remaining step is purely temporal integration, leveraging the meshfree nature of SPH.

\bibliography{biblio}

\end{document}